\documentclass[11pt,twoside]{book}
\usepackage{konkolyproc2}
\usepackage{longtable}
\usepackage{amsmath,amssymb}
\usepackage{graphicx}
\usepackage{lscape}
\usepackage{index}
\usepackage{natbib}
\usepackage{bigdelim}
\usepackage{multirow}
\makeindex

\begin{document}

%%%%%%%%%%%%%%%%%%%%%%%%%%%%%%%%%%%%%%%%%%%%%%%%%%%%%%%%%%%%%%%%%%%%%%%%%%
\pagestyle{myheadings}
\setcounter{equation}{0}\setcounter{figure}{0}\setcounter{footnote}{0}
\setcounter{section}{0}\setcounter{table}{0}\setcounter{page}{1}
\markboth{Plachy et~al.}{Target selection of pulsating variables for space-based photometry}
\title{Target selection of classical pulsating variables for space-based photometry}
\author{E. Plachy$^1$, L. Moln\'ar$^1$, R. Szab\'o$^1$, 
K. Kolenberg$^{2,3,4}$, \& E. B\'anyai$^5$}
\affil{$^1$Konkoly Observatory, Research Centre for Astronomy and Earth Sciences, 
Hungarian Academy of Sciences, H-1121, Budapest, Konkoly Thege Mikl\'os \'ut 15-17, Hungary\\
$^2$Instituut voor Sterrenkunde, Leuven, Belgium\\
$^3$Harvard-Smithsonian Center for Astrophysics, Cambridge, USA\\
$^4$University of Antwerp,  Antwerp, Belgium\\
$^5$E\"otv\"os Lor\'and University, Budapest, Hungary\\}
%%%%%%%%%%%%%%%%%%%%%%%%%%%%%%%%%%%%%%%%%%%%%%%%%%%%%%%%%%%%%%%%%%%%%%%%%%%

\begin{abstract}
In a few years the \textit{Kepler} and \textit{TESS} missions will provide 
ultra-precise photometry  for thousands of RR Lyrae and hundreds of Cepheid stars. 
In the extended \textit{Kepler} mission all targets are proposed in the Guest 
Observer (GO) Program, while the \textit{TESS} space telescope will work with full 
frame images and a \linebreak  $\sim$15-16th mag brightness limit with the 
possibility of short cadence measurements for a limited number of pre-selected 
objects. This paper highlights some details of the enormous and important work 
of the target selection process made by the members of Working Group 7 (WG\#7) 
of the \textit{Kepler} and \textit{TESS} Asteroseismic Science Consortium.   
\end{abstract}

\vspace*{-4mm}
\section{\textit{K2} target selection and proposals}
\vspace*{-2mm}
The new era of space-based photometry has already begun. The reaction wheel 
failure of \textit{Kepler} space telescope opened a great possibility to 
build up a golden sample for many types of variable stars. In the \textit{K2} 
mission, \textit{Kepler} observes the ecliptic plane and changes field of view 
in every $\sim$80 days \citep{K2}. The mission started in March of 2014 with 
Campaign 0 (C0) and is planned to end in April of 2018 with Campaign 17 (C17). 
The invitation to the scientific community to propose targets for 
ultra-precise measurements in the GO Program motivates many astronomers 
to come forward with new ideas. Working Group 7 (WG\#7) is interested in 
RR~Lyrae and Cepheid stars, and it is responsible for the proposals of these 
objects for each campaign. 

\begin{table}[!ht] 
%\tiny
\scriptsize
%\footnotesize
\caption{RR Lyrae and Cepheid proposals in the {\it K2} mission.}
\smallskip
\begin{center}

\begin{tabular}{llll}
\tableline
\noalign {\smallskip} 
Campaign & Proposal Number & PI & Topic  \\ 
\noalign{\smallskip}
\tableline
\noalign{\smallskip}
C0&GO0051 & Moln\'ar & Long cadence Cepheid targets\\ 
&GO0053 & Plachy & Short cadence Cepheid targets\\ 
&GO0055 & Szab\'o & Long cadence RR Lyrae targets \\ 
&GO0124 & Kolenberg & Short cadence RR Lyrae targets \\ 
\noalign{\smallskip}
\tableline
\noalign{\smallskip}
C1&GO1018 & Plachy & Long cadence RR Lyrae targets\\ 
&GO1019& Moln\'ar & RR Lyrae in the dwarf galaxy Leo IV\\ 
&GO1021 & Moln\'ar & Long cadence Cepheid targets\\ 
&GO1067 & Kolenberg & Short cadence RR Lyrae targets\\ 
\noalign{\smallskip}
\tableline
\noalign{\smallskip}

C2 \& C3 &GO2027 $\&$ GO3027 & Plachy & Short cadence RR Lyrae targets\\ 
&GO2039& Moln\'ar & Pulsating variables in M4 and M80\\ 
&GO2040 $\&$ GO3040 & Moln\'ar & Long cadence RR Lyrae targets\\ 
&GO2041 $\&$ GO3041 & Moln\'ar & Type I and II Cepheids\\ 
\noalign{\smallskip}
\tableline
\noalign{\smallskip}
C4 \& C5 &GO4066 $\&$ GO5066 & Moln\'ar &  Type I and II Cepheids\\ 
&GO4069 $\&$ GO5069 & Szab\'o & Exploiting RR Lyrae stars\\ 
\noalign{\smallskip}
\tableline
\noalign{\smallskip}
C6 \& C7&GO6082 $\&$ GO7082 & Kolenberg  & RR Lyrae stars from different populations \\ 
&GO7014 & Moln\'ar & Sampling the Cepheid instability strip\\ 
 \noalign{\smallskip}
\tableline
\noalign{\smallskip}
C8 \& C10&GO3-0039 & Szab\'o  &  Pulsation dynamics and \\
& & & Galactic structure of RR Lyrae stars \\ 
&GO3-0041 & Moln\'ar & Extragalactic Cepheids in IC 1613\\ 
 \noalign{\smallskip}
\tableline
\noalign{\smallskip}
C9& DDT & Smolec  &  RR Lyrae stars in the Galactic bulge\\
 &DDT & Plachy & Classical and Type II Cepheids in the Bulge\\ 

\noalign{\smallskip}
\tableline
\noalign{\smallskip}

C11, C12 \& C13 &GO4-0070 & Plachy  &  Cepheids throughout the Galaxy\\
  &GO4-0111 & Moln\'ar  &   The grand \textit{K2} RR Lyrae survey\\
\noalign{\smallskip}
\tableline
\noalign{\smallskip}
\end{tabular}  \end{center}
 \normalsize
\end{table}

WG\#7 has submitted altogether 22 proposals, listed in Table~1, 
at the time of writing this article. According to the initial concept, 
proposals were separated by variability types and cadence type (30 or 
1 min). We dedicated proposals to dwarf galaxies and globular clusters as well. 
After C1, joint proposals were submitted for two or three fields. We typically 
submitted four proposals for each of the first four campaigns. Afterwards the 
calls were made through the 2-step process of the NASA proposal system. 
Since then we have submitted united proposals for short and long cadence targets 
for the RR Lyrae and Cepheid stars.

The main goal of the proposals is to obtain all RR Lyrae and Cepheid targets 
that fall on the {\it K2} fields. To build up a golden sample it is crucial to 
calibrate the classification and analysis methods. Our scientific 
justifications focus on the most pressing questions raised recently: 
the origin of dynamical phenomena, the low amplitude additional modes 
and the mysterious period ratios. The existence of nonradial modes and 
the explanation of the Blazhko effect are still open questions. The {\it K2} 
mission also provides the opportunity for population and Galactic structure 
studies as well as statistical analysis of various phenomena. 
A limited number of targets is observed with 1 minute sampling. 
We select the most interesting or rare type of targets to propose for 
short cadence mode \citep{IBVS}. 

Our target selection process was first used for the Two-Wheel Concept 
Engineering Test that led to a detailed analysis of 33 RR Lyrae stars 
\citep{E2}. First we collect all known RR Lyrae and Cepheid candidates from 
the SIMBAD and VSX databases. Several sky surveys provide semi-automated 
variability catalogues and downloadable light curves as well. We found the 
Catalina Sky Survey \citep{CSS},  Lincoln Near Earth Asteroid Research 
\citep{LINEAR}, All Sky Automated Survey \citep{ASAS}, Northern Sky Variability 
Survey \citep{NSVS} extremely useful for the target selection.
The next step is to select the stars that fall on silicon. We use the 
{\tt K2FoV\footnote{http://keplerscience.arc.nasa.gov/software.html, 
https://github.com/KeplerGO/K2fov}} tool that has been developed by the Kepler 
GO Office for  this purpose. The sky surveys overlap, so we have to do the 
cross-identification of the different catalogues. Because of the relatively 
high uncertainty in the coordinates of certain objects, we found this step 
to be more reliable if done manually. The last and the most important step 
is to check of the folded light curves of the targets. The visual 
inspection can reveal misclassified objects, erroneous published periods 
and potential short cadence targets as well. 
%We propose the interesting RR~Lyraes for short cadence mode, 
%%%which are mostly the rare double-mode RRd stars, the confirmed first overtone 
%%%RRc stars, the extremely modulated and unusually long- or short-period 
%%%fundamental-mode RRab stars.

\begin{table}[!ht] 
%\tiny
\scriptsize
%\footnotesize
\caption{Number of proposed and accepted RR Lyrae and Cepheid targets.}
\smallskip
\begin{center}
\begin{tabular}{ccccccc}
\tableline
\noalign {\smallskip} 
 Campaign &
      \multicolumn{3}{c}{RR Lyrae targets} &
      \multicolumn{3}{c}{Cepheid targets}  \\
 & Proposed & Accepted & Success rate & Proposed & Accepted & Success rate\\ 
\noalign{\smallskip}
\tableline
\noalign{\smallskip}
C0 & 68 & 10 &\textbf{$\sim$15\%}& 54 &13 &\textbf{$\sim$24\%}\\ 
C1 & 136 & 18 &\textbf{$\sim$13\%}& 4 &4 &\textbf{100\%}\\ 
C2 & 86 & 61 &\textbf{$\sim$71\%}& 8 &8 &\textbf{100\%}\\ 
C3 & 117 & 82 &\textbf{$\sim$70\%}& 1 &1 &\textbf{100\%}\\ 
C4 & 86 & 83 &\textbf{$\sim$97\%}& 7 &7 &\textbf{100\%}\\ 
C5 & 89 & 88 &\textbf{$\sim$99\%}& 4 &4 &\textbf{100\%}\\ 
C6 & 206 & 206 &\textbf{100\%}& - & - & -\\ 
C7 & 528 & 528 &\textbf{100\%}& 10 &9 &\textbf{90\%}\\
C8 & 85 & 85 &\textbf{100\%}& 190 &190&\textbf{100\%}\\ 
C9 & 200 & N/A &N/A& 184 &N/A&N/A\\
C10 & 224 & N/A &N/A& 2 &N/A&N/A\\ 
C11 & 1629 & N/A &N/A& 164 &N/A&N/A\\
C12 & 181 & N/A &N/A& 6 &N/A&N/A\\
C13 & 94 & N/A &N/A& 10 &N/A&N/A\\

\noalign{\smallskip}
\tableline
\end{tabular}  
\end{center}
 \normalsize
\end{table}

Table 2 summarizes the number of targets proposed and accepted in each 
field so far. The success rate of the proposals is quite high. 
The reason for the initial low percentages is mostly technical: the lack 
of {\tt K2FoV} tool in C0 or the 5 degree roll of the field of view in C3.  
In some cases targets fell near the edge of the CCD and in one case the 
target is too bright to be measured.  
% A few targets of C2 are in the superstamp of M4,that did not receive 
%%%EPIC (Ecliptic Plane Input Catalogue) numbers, but data can be extracted 
%%%with difference image analysis. 
We note that the number of RR Lyrae and Cepheid stars measured in {\it K2} 
are not identical to the numbers of Table~2. Several additional targets are 
located in the superstamps of the globular clusters (C2) and the Galactic 
bulge (C9). Moreover, we predict a significant fraction of misclassified 
objects among the RRc, RRd and Cepheid candidates, but we also expect new 
findings among the pre-classified binaries. 
%{\it K2} fields are fixed until C13 and proposed until C17.
   
\section{\textit{TESS} target selection}

%C9 will observe the Galactic bulge to monitor microlensing events: 
%%%this dense region contains thousands of RR Lyraes and hundreds of Cepheids. 
%%%DDT proposals can be submitted for C9 -- our plan here is to focus on 
%%%binary RR Lyrae candidates and the RRc stars that are all suspected to 
%%%contain the mysterious $\sim$0.61 period ratio \citep{netzel}. Type II Cepheids 
%%%also populate this region, and this type needs the ultra-precise photometry 
%%%the most since there are only a few candidates observed from space so far. 

$TESS$ will observe almost the entire sky and will download full frame 
images with 30-minute cadence \citep{TESS}. Few hundred thousand targets 
will be selected for 2-minute sampling. Most of these are exoplanet candidates 
but 5 percent will be devoted to asteroseismic targets proposed by the \textit{TESS} 
Asteroseismic Science Consortium. The target selection of these objects need 
the same careful process that we use in {\it K2} mission. The major difference 
will be in the brightness limit that is expected to be $\sim$12th mag for the 
short cadence objects reducing the number of the potential targets. In Figure~1 
we plotted the continuous viewing zones of \textit{TESS} around the ecliptic poles. 
8 (18) RR~Lyrae and 3 (35) Cepheid short cadence candidates around the North 
(South) Ecliptic Poles are marked with black circles and diamonds, respectively.   
   
\begin{figure}[!ht]
\includegraphics[width=1.0\textwidth]{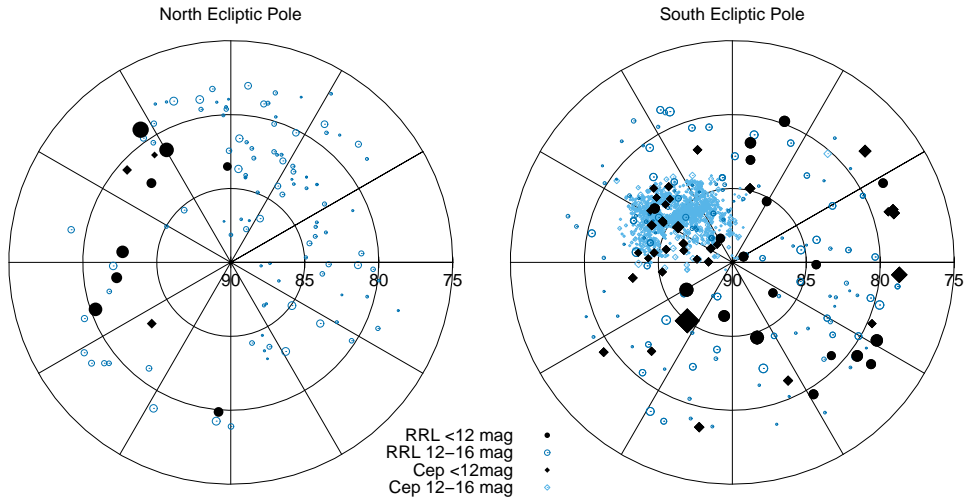}
\caption{RR Lyrae and Cepheid stars in the \textit{TESS} continuous viewing zone.} 
\label{authorsurname-fig1} 
\end{figure}   

\section*{Acknowledgements}
This research has been supported by the LP2014-17 Program of the Hungarian 
Academy of Sciences, by the NKFIH K-115709, the OTKA NN-114560 and 
the PD-116175 grants of the Hungarian National Research, Development and 
Innovation Office. The research leading to these results has received funding 
from the European Community's Seventh Framework Programme (FP7/2007-2013) 
under grant agreements no. 269194 (IRSES/ASK), no. 312844 (SPACEINN) and 
ESA PECS Contract No. 4000110889/14/NL/NDe. K.K. is grateful for the support 
of Marie Curie IOF grant 255267 SASRRL (FP7). L.M.\ was supported by the 
J\'anos Bolyai Research Scholarship of the Hungarian Academy of Sciences. 
This research has made use of the SIMBAD database, operated at CDS,
Strasbourg, France, and the International Variable Star Index (VSX) database, 
operated at AAVSO, Cambridge, Massachusetts, USA.

\end{document}